\documentclass[electronic]{vgtc}                          

\usepackage[utf8]{inputenc}

\usepackage[usenames,dvipsnames,svgnames,table,x11names]{xcolor}
\usepackage[breaklinks]{hyperref}
\usepackage{mathptmx}
\usepackage{graphicx}
\usepackage{subcaption}
\usepackage{hyphenat}
\usepackage{booktabs}
\usepackage{pifont}
\usepackage{listings} 
\usepackage{colortbl}
\usepackage{xfrac}
\usepackage[inline]{enumitem}
\usepackage{amsmath}
\usepackage{gensymb}
\usepackage{multirow}
\usepackage[thinlines]{easytable}
\usepackage{algorithm}
\usepackage{times}
\usepackage{url}
\usepackage{blindtext}
\usepackage{wasysym}
\usepackage{color}
\DeclareUnicodeCharacter{00A0}{ }
\definecolor{DarkBlue}{rgb}{0,0,0.1}
\hypersetup{
  colorlinks   = true,
  citecolor    = blue,
  linkcolor		 = blue,
  filecolor		 = blue,      
  urlcolor		 = blue,
  linkbordercolor = TealBlue,
  citebordercolor = TealBlue,
  urlbordercolor = TealBlue,
}
\onlineid{0}

\vgtccategory{Research}

\title{Procedural Planetary Multi-resolution Terrain Generation for Games}

\author{Ricardo B. D. d'Oliveira\thanks{e-mail: ricardo.barros(at)ufba.br}
  \and Antonio L. Apolinário Jr.\thanks{e-mail: antonio.apolinario(at)ufba.br}}
\affiliation{\scriptsize Federal University of Bahia, Brazil}

\teaser{
  \begin{subfigure}[b]{0.09\textwidth}
    \includegraphics[width=\textwidth]{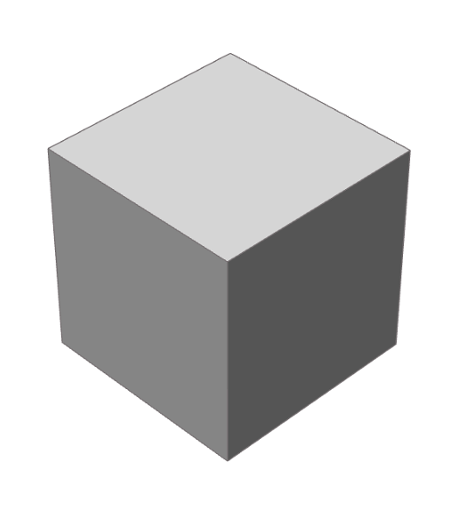}
    \label{fig:myproposal00}
  \end{subfigure}  
  \begin{subfigure}[b]{0.09\textwidth}
    \raisebox{3\height}{$\xrightarrow[\scriptscriptstyle\text{control mesh }]{\scriptscriptstyle\text{create \& subdivide}}$}
  \end{subfigure} 
  \begin{subfigure}[b]{0.09\textwidth}
    \includegraphics[width=\textwidth]{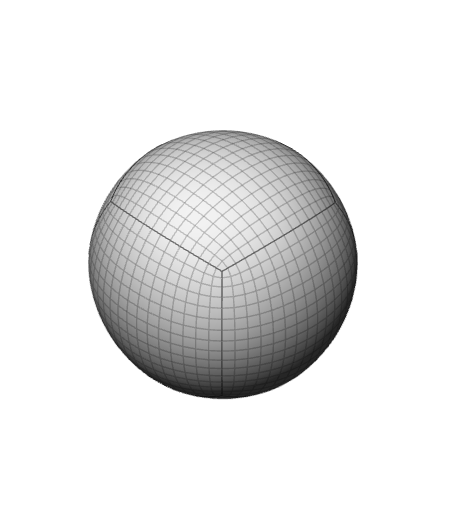}
    \label{fig:myproposal01}
  \end{subfigure}  
  \begin{subfigure}[b]{0.055\textwidth}
    \raisebox{3\height}{$\xrightarrow[\scriptscriptstyle\text{heightmap}]{\scriptscriptstyle\text{generate}}$}
  \end{subfigure} 
  ~
  \begin{subfigure}[b]{0.08\textwidth}
    \includegraphics[width=\textwidth]{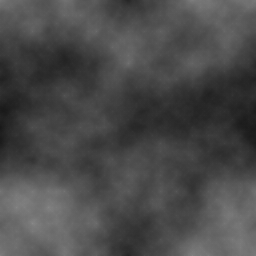}
    \label{fig:terrain01}
  \end{subfigure}  
  ~
  \begin{subfigure}[b]{0.07\textwidth}
    \raisebox{3\height}{$\xrightarrow[\scriptscriptstyle\text{control mesh}]{\scriptscriptstyle\text{displace}}$}
  \end{subfigure} 
  \begin{subfigure}[b]{0.09\textwidth}
    \includegraphics[width=\textwidth]{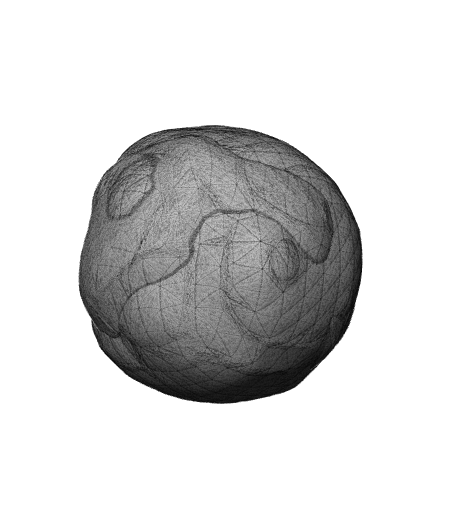}
    \label{fig:myproposal02}
  \end{subfigure}  
  
  \begin{subfigure}[b]{0.055\textwidth}
    \raisebox{3\height}{$\xrightarrow[\scriptscriptstyle\text{tessellate}]{\scriptscriptstyle\text{zoom 1x}}$}
  \end{subfigure} 
  ~
  \begin{subfigure}[b]{0.09\textwidth}
    \includegraphics[width=\textwidth]{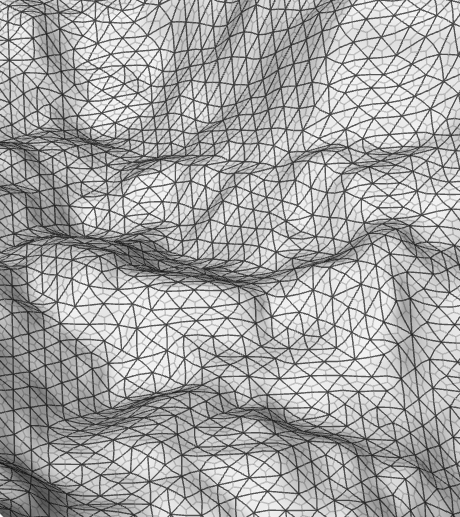}
    \label{fig:myproposal03}
  \end{subfigure}  
  ~
  \begin{subfigure}[b]{0.1\textwidth}
    \raisebox{3\height}{$\xrightarrow[\scriptscriptstyle\text{tessellate further}]{\scriptscriptstyle\text{zoom 2x}}$}
  \end{subfigure} 
  ~
  \begin{subfigure}[b]{0.09\textwidth}
    \includegraphics[width=\textwidth]{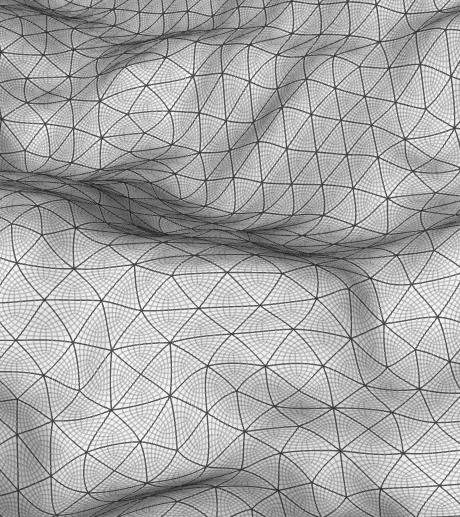}
    \label{fig:myproposal04}
  \end{subfigure}  
  ~
  \begin{subfigure}[b]{0.1\textwidth}
    \raisebox{3\height}{$\xrightarrow[\scriptscriptstyle\text{details with noise}]{\scriptscriptstyle\text{zoom to surface}}$}
  \end{subfigure} 
  ~
  \begin{subfigure}[b]{0.09\textwidth}
    \includegraphics[width=\textwidth]{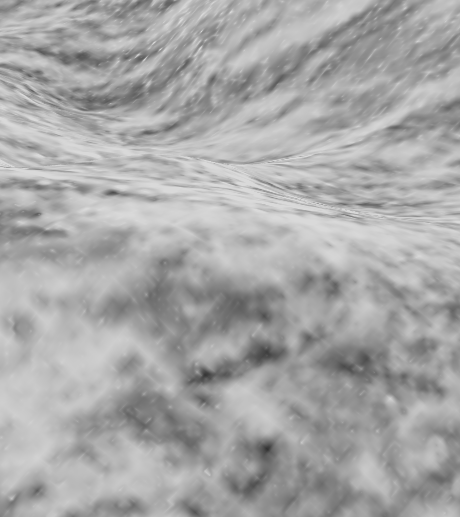}
    \label{fig:myproposal05}
  \end{subfigure}
  \caption{Our proposal's pipeline for procedurally generating multi-resolution terrains.}
  \label{fig:teaser}
}

\abstract{
	Terrains are the main part of an electronic game. To reduce human effort on game development, procedural techniques are used to generate synthetic terrains. However rendering a terrain is not a trivial task. Their rendering techniques must be optimal for gaming. Specially planetary terrains, which must account for precision and scale conversion. Multi-resolution models are best fit to planetary terrains. An observer can change his point of view without noticing any decrease in visual quality. There are several proposals regarding real-time terrain rendering with multi-resolution models, and there are game engines capable of generating large scale terrains with fixed resolution. However for the best of our knowledge, it was noticed that there are no techniques which combine both aspects. In this paper we present a new technique capable of generating large-scale multi-resolution terrains, whichcan be rendered and viewed at different scales. Rendering large scale models with high definition and low scale areas with finer details added with the aid of procedural content generation. 
  \smallskip
  
} 

\begin{document}
  
  
  \firstsection{Introduction}
  
  \maketitle    
  Computer gaming is increasingly present in our lives. Its development is a challenging task, considering its complexity and its assets \cite{bergeron2016}. Its assets can be comprised of graphical objects which usually have high definition models \cite{bergeron2016}. Manual modeling can be time consuming depending on the desired quality \cite{bergeron2016}. Procedural content generation can be an alternative to manual content creation, it can randomize its output and reduce manual intervention \cite{hendrikx2013}. Games benefit directly from procedural content generation, because textures, audios and 3D models can be generated procedurally, with little to no human intervention \cite{hendrikx2013}.

  PCG is widespread used in games, titles such as \textit{Spore} \cite{compton2007}, \textit{Torchlight} \cite{campbell2010}, \textit{Diablo 3} \cite{lambe2012} and \textit{No Man's Sky} \cite{lambe2012} use PCG to create 3D models, terrains, textures, scenarios, providing a unique experience in each playthrough. PCG removes the burden off game development, reducing its costs and its development time. The earliest use of PCG in eletronic games dates back to \textit{Elite}'s approach on generating an universe \cite{braben84}, while the most outstanding can be considered the game \textit{.kkrieger} \cite{kkrieger2006} which is completely developed using PCG, grainy textures are created using Perlin Noise \cite{perlin85} and stored in memory, the terrain is deformed using midpoint displacement \cite{fournier82} and meshes are created from basic forms and then deformed to a specific shape, resulting in a 97,280 bytes executable.
  \begin{figure}[!htb]
    \centering
    \includegraphics[width=.48\textwidth]{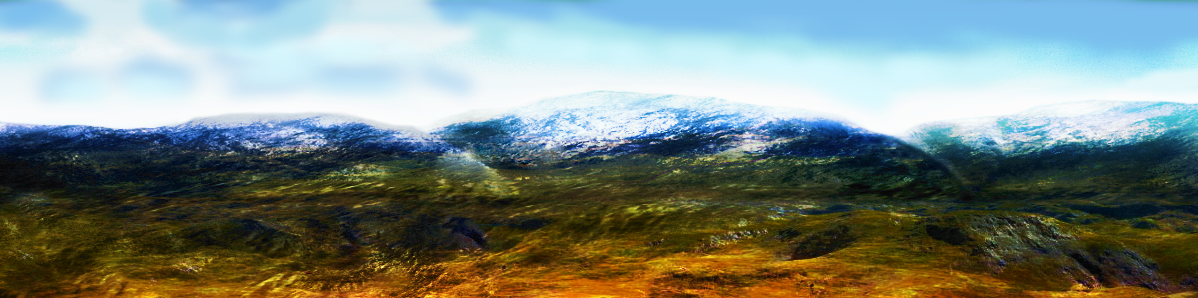}
    \caption{Procedurally generated terrain using Blender.}
    \label{fig:terrain}
  \end{figure}
  
  Virtual terrains, are game assets that, can be considered the main part of a game, because it is on the terrain where the game takes place \cite{bergeron2016}. A procedural method can be used to generate a terrain with a high visual quality, such as depicted in Figure \ref{fig:terrain}. PCG can be used to generate vast \cite{oneil2006b} or infinite terrains \cite{parberry2014}. O'Neil \cite{oneil2006b} adapted the ROAM terrain rendering algorithm \cite{duchaineau97} to tessellate a cube into an oblate sphere, applying a fractal algorithm to its 
  surface, thus generating a planet. However O'Neil \cite{oneil2006b} demonstrated that using CPU terrain rendering techniques is not efficient, performance improvement could be achieved by using dedicated GPU techniques. Also, large terrains should be carefully dealt with to avoid precision errors \cite{kooima2009}. 

  A virtual terrain can be generated and rendered in different scale levels, there are several proposals regarding real-time terrain rendering with multi-resolution models, and there are game engines capable of generating large scale terrains with fixed resolution. However for the best of our knowledge, it was noticed that there are no techniques which combine both aspects. In this paper we present a new technique capable of generating large-scale multi-resolution terrains, which can be rendered and viewed at different scales, and at the lowest scale details are going to be added with PCG.
  
  \section{Basic Concepts}
  DEM is a shorthand for Digital Elevation Model. It can be a textual or binary data representing a set of elevation samples \cite{maune2001}. Which in turn can be vector, such as Triangulated Irregular Grid -- TIN, or raster. A raster DEM is known as heightmap, such a heightmap can be a greyscale image (Figure \ref{fig:terrainGPU01}). Where each pixel represents the elevation data if mapped correctly to a 3D grid mesh.
  
  Procedural content generation -- PCG can be described as a serie of techniques that enable automatic content generation, random textures, 3d models and even audio can be created using PCG.
  
  The Programmable Graphics Pipeline -- PGP is the way to manipulate how a 3d scene is going to be rendered by coding each step of the graphics pipeline through shaders, shaders in turn are pieces of code than run directly into the GPU \cite{ginsburg2014}. In this proposal we are discussing mainly both the Tessellation Shader -- TS, which allows to refine simpler meshes into finer ones, and the Fragment Shader -- FS, which is used to apply lighting, color blending, process and render a fragment, for example a pixel.
  
  \section{Related Work}
  To represent scenes with large-scale models a vast amount of computational resources, such as RAM, CPU and GPU, must be present. Our approach tends toward gaming, thus those resources must be managed properly. There are two distinct approaches to manage these resources, either by using level of detail (LOD), specifically terrain rendering LOD techniques \cite{strugar2009}, or load balancing, such as CPU-GPU coupled computation techniques \cite{bernhardt2011}. LOD techniques were originally made for CPU, but as GPU became sophisticated those techniques were implemented with TS.
  
\begin{figure}[!htb]
    \centering
    \begin{subfigure}[b]{0.105\textwidth}
      \includegraphics[width=\textwidth]{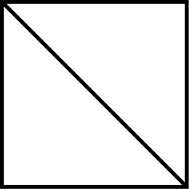}
      \caption{}
      \label{fig:bintree0}
    \end{subfigure}  
    ~
    \begin{subfigure}[b]{0.105\textwidth}
      \includegraphics[width=\textwidth]{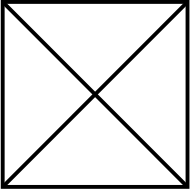}
      \caption{}
      \label{fig:bintree1}
    \end{subfigure}  
    ~
    \begin{subfigure}[b]{0.105\textwidth}
      \includegraphics[width=\textwidth]{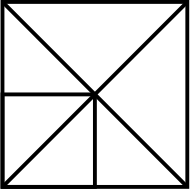}
      \caption{}
      \label{fig:bintree2}
    \end{subfigure}     
    ~
    \begin{subfigure}[b]{0.105\textwidth}
      \includegraphics[width=\textwidth]{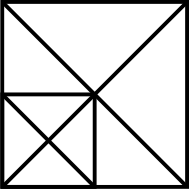}
      \caption{}
      \label{fig:bintree3}
    \end{subfigure}      
    
    \begin{subfigure}[b]{0.105\textwidth}
      \includegraphics[width=\textwidth]{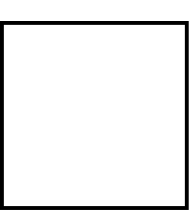}
      \caption{}
      \label{fig:quadtree0}
    \end{subfigure}  
    ~
    \begin{subfigure}[b]{0.105\textwidth}
      \includegraphics[width=\textwidth]{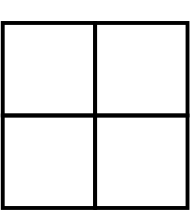}
      \caption{}
      \label{fig:quadtree1}
    \end{subfigure}  
    ~
    \begin{subfigure}[b]{0.105\textwidth}
      \includegraphics[width=\textwidth]{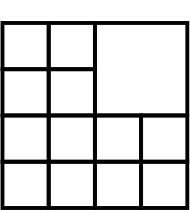}
      \caption{}
      \label{fig:quadtree2}
    \end{subfigure}     
    ~
    \begin{subfigure}[b]{0.105\textwidth}
      \includegraphics[width=\textwidth]{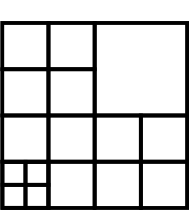}
      \caption{}
      \label{fig:quadtree3}
    \end{subfigure}   
    \caption{Adaptative meshes with control data structure, the upper part uses hierarchical bintrees, the lower part uses quadtrees.}
    \label{fig:dataStructures}
  \end{figure}

  Terrain rendering is a well known research field among computer graphics, techniques such as ROAM \cite{duchaineau97} and BDAM \cite{cignoni2003a} are prime examples of CPU terrain rendering with LOD, both use hierarchical binary trees (Figure \ref{fig:bintree0}) to represent the terrain LOD. In contrast, more recent techniques such as \cite{strugar2009}, \cite{dick2009} and \cite{valdetaro2010} have a GPU oriented approach, using quadtrees (Figure \ref{fig:quadtree0}) to represent the terrain LOD and using the PGP.

  Terrain rendering algorithms can usually deal with vast terrains. But when it comes to planetary rendering adaptions must be made. In Cignoni et al. \cite{cignoni2003b} the BDAM algorithm \cite{cignoni2003a} is adapted and coupled with a out-of-core technique to deal with large data and textures, being capable of rendering massive objects such as the planet Mars. O'Neil \cite{oneil2006b} is an adaptation of the ROAM algorithm is made to procedurally generating planets. That approach consists of creating six terrains and interpolating them through a center of mass, generating a planetary body. Kooima et al. \cite{kooima2009} proposes a GPU centred approach, employing general purpose GPU programming using little to no pre-processing, to render planetary terrains. 
  
  Multi-resolution terrain rendering is accomplished by generating different mesh resolutions for each field of view, for example when viewing from the ground a terrain has a different resolution than viewed from space. While some LOD techniques can be applied to generate different terrain resolutions, finer details cannot be represented or recovered. There are some approaches to circumvent this by either using TS \cite{kai2006} or by using multiple DEM \cite{danielson2011}.
  
  Pajarola and Gobbetti \cite{pajarola2007} surveys on multi-resolution models for terrain rendering, comparing large scale terrain visualization techniques and their data structures. Pointing out that even though some of these techniques are simple to implement, they do not provide full support on adaptive surfaces. Lindstrom and Cohen \cite{Lindstrom2010} proposes a on-the-fly decompression and rendering technique for multi-resolution where geometry is encoded through compression to be decompressed on the CPU or streamed for parallel decoding on the GPU. 
  
  \begin{figure}[!htb]
    \centering
    \begin{subfigure}[b]{0.09\textwidth}
      \includegraphics[width=\textwidth]{terrain01}
      \caption{A heightmap}
      \label{fig:terrainGPU01}
    \end{subfigure}  
    \begin{subfigure}[b]{0.055\textwidth}
      \raisebox{3\height}{$\xrightarrow[\scriptscriptstyle\text{base mesh}]{\scriptscriptstyle\text{create}}$}
    \end{subfigure}  
    \begin{subfigure}[b]{0.13\textwidth}
      \includegraphics[width=\textwidth]{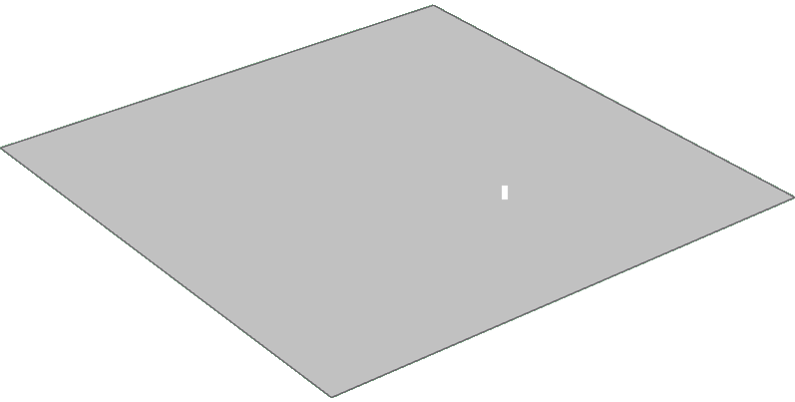}
      \label{fig:terrainGPU02}
    \end{subfigure}   
    \begin{subfigure}[b]{0.04\textwidth}
      \raisebox{3\height}{$\xrightarrow[\scriptscriptstyle\text{quadtree}]{\scriptscriptstyle\text{create}}$}
    \end{subfigure} 
    \begin{subfigure}[b]{0.13\textwidth}
      \includegraphics[width=\textwidth]{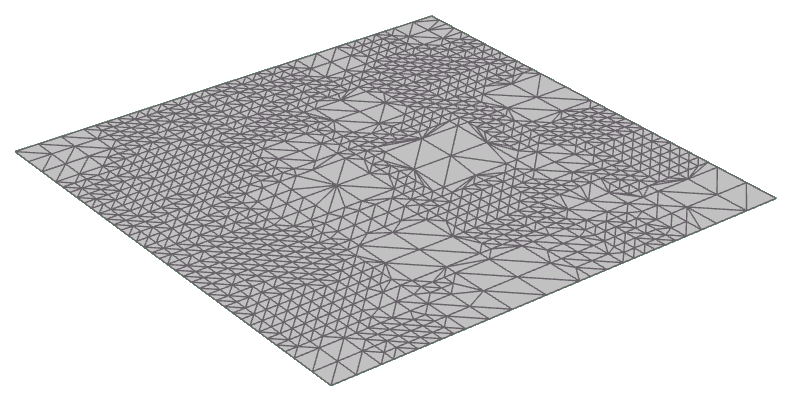}
      \label{fig:terrainGPU03}
    \end{subfigure}
    
    \begin{subfigure}[b]{0.35\textwidth}
      \includegraphics[width=\textwidth]{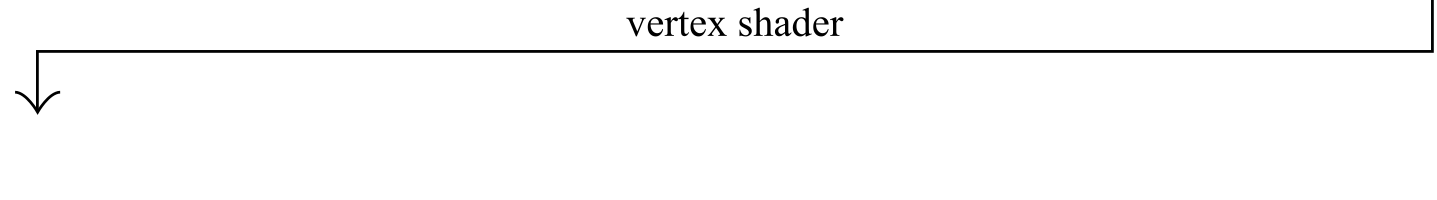}
    \end{subfigure}
    
    \begin{subfigure}[b]{0.12\textwidth}
      \includegraphics[width=\textwidth]{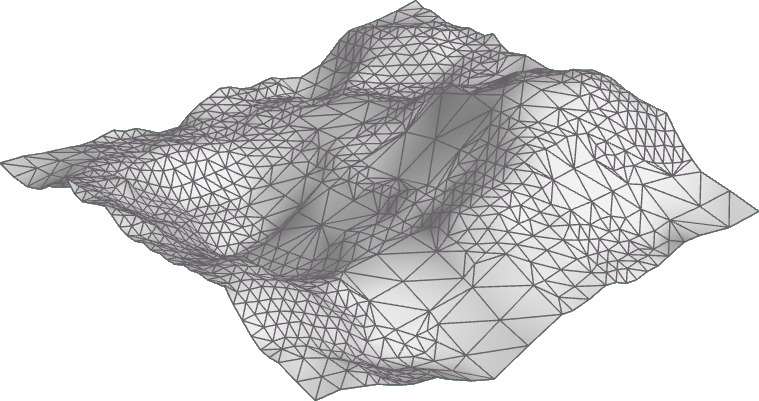}
      \label{fig:terrainGPU04}
    \end{subfigure}  
    \begin{subfigure}[b]{0.05\textwidth}
      \raisebox{3\height}{$\xrightarrow[\scriptscriptstyle\text{shader}]{\scriptscriptstyle\text{tessellation}}$}
    \end{subfigure} 
    \begin{subfigure}[b]{0.12\textwidth}
      \includegraphics[width=\textwidth]{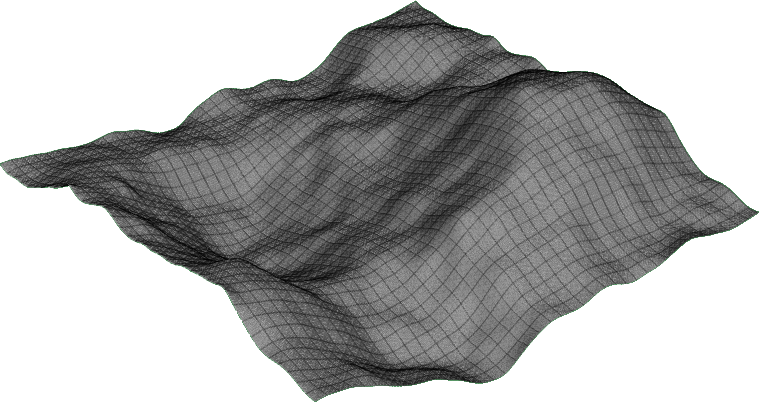}
      \label{fig:terrainGPU05}
    \end{subfigure}   
    \begin{subfigure}[b]{0.04\textwidth}
      \raisebox{3\height}{$\xrightarrow[\scriptscriptstyle\text{shader}]{\scriptscriptstyle\text{fragment}}$}
    \end{subfigure} 
    \begin{subfigure}[b]{0.12\textwidth}
      \includegraphics[width=\textwidth]{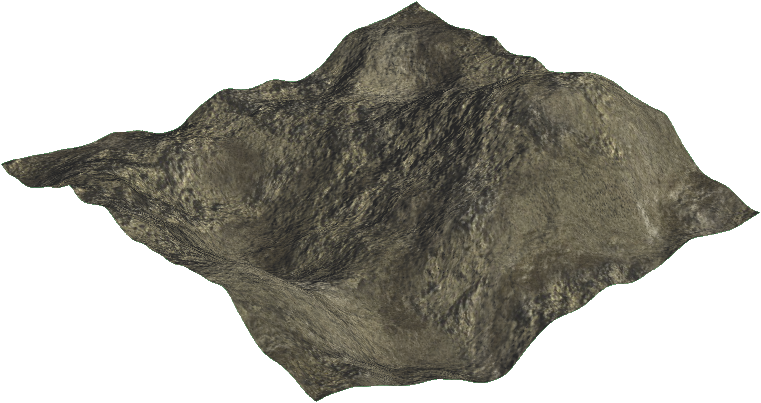}
      \label{fig:terrainGPU06}
    \end{subfigure}
    \caption{Terrain rendering graphics pipeline, the upper part runs on CPU while the lower runs on GPU.}
    \label{fig:terrainGPU}
  \end{figure}
  
  Generally speaking, current terrain rendering techniques use PGP, and their pipeline is similar to the one shown in the Figure \ref{fig:terrainGPU}. Where a heightmap is read and stored as a texture, the quadtree is built following bottom-up strategy. In such way it can be decided that each leaf node has the highest possible LOD. Each node of the quadtree represents a patch. Each patch is passed to the Vertex Shader to be displaced accordingly to its respective coordinates in relation to the heightmap. In the TS the geometry can be refined depending on the desired LOD, while in the FS lightning and color blending can be applied. 
  
  Terrain rendering techniques which use TS are more applicable in the context of this proposal, they are fair better at generating global features for a terrain \cite{kang2015}. Using TS a grid mesh can be smoothed, making it easier to add details, thus we can also explore LOD and multi-resolution approaches. 
  
\begin{figure}[!htb]
    \centering
    \begin{subfigure}[b]{0.23\textwidth}
      \includegraphics[width=\textwidth]{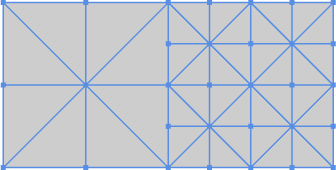}
      \caption{Mesh with two LOD patches.}
      \label{fig:quadFix0}
    \end{subfigure}  
    ~
    \begin{subfigure}[b]{0.23\textwidth}
      \includegraphics[width=\textwidth]{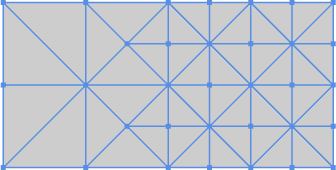}
      \caption{Fixed mesh without cracks.}
      \label{fig:quadFix1}
    \end{subfigure} 
    \caption{To solve a crack after tesselating, a midpoint is added on the edges which are adjacent to the patch of higher definition.}
    \label{fig:quadFix}
  \end{figure}

  While tessellating a mesh two issues can arise. If a patch is tessellated using a factor and it's adjacent patch part isn't tessellated or uses a lesser factor inconsistencies can appear. These inconsistencies are known as cracks (Figure \ref{fig:quadFix0}). The other issue revolves around the way LOD is managed, if LOD isn't chosen properly or navigation speed is too fast visual inconsistencies can happen. Both Kim and Baek \cite{kim2014} and Song et al. \cite{ge2017} propose techniques for terrain rendering using TS.
  
  To resolve cracks,  Kim and Baek \cite{kim2014} uses quadilateral patches (Figure \ref{fig:quadFix0}). Where both edges of the lower definition area, which are adjacent to the higher definition one, are bisected along with its largest edge, resulting in a mesh without any inconsistency left as shown in Figure \ref{fig:quadFix1}. Song et al. \cite{ge2017} algorithm uses tessellation control points, where its algorithm creates as many as needed points to represent a basic structure of geometry clipmaps. Instead of creating all vertices for a primitive, such as subdividing a base mesh with triangles.  
  
  With a TS technique terrains of varying quality can be rendered, and if coupled with procedural terrain generation, not only a terrain can be generated, but details can be added. This is where TS come in, they provide ways to manage LOD and we can decide in which LOD finer details can be further added. Procedural Generation is usually correlated with fractal algorithms, \cite{perlin85} and \cite{fournier82} are the first proposal for terrain generation through fractals, and both argue that fractals are well suit for that task because they mimic natural events and its stochastic behavior. Fractal algorithms are usually described as noise algorithms \cite{fournier82}\cite{perlin85}\cite{lagae2010}. While these techniques are relatively old, they are still used \cite{blatz2017}, because of their efficiency and simplicity. Perlin noise can generate content with natural looking and can be used to generate virtually any kind of media ranging from 2d objects such as textures and audios to 3d models \cite{blatz2017}. Perlin Noise is not restricted to content generation but it can be use for generic mesh refinement \cite{boubekeur2005}. There are several Perlin noise implementations \cite{lagae2010}, among them there is \cite{haase2012}, an implementation for generating fractal noises on the GPU, which is applicable to the context of this proposal.
  
  While Terrain with TS is well suit for global visualization. For instance, \cite{tiju2012} is an implementation that generates terrain with Perlin Noise, then the mesh output is tessellated according to the viewer distance, giving a terrain with high definition if the camera is close to the surface, and a rough and less detailed as a viewer looks into distance, this implementation follows a pipeline similar to the one depicted in Figure \ref{fig:terrainGPU}.
  
  Whereas PCG is adequated for adding finer details to a mesh, such as skin harshness, surface porosity or rugosity, among others. For example, a fractal noise can be generated on the FS, it then can be applied as a displacement map, creating a rough aspect on the mesh. This approach is used on \cite{bangay2017}. It proposes a deterministic tiling strategy with fractal noise to achieve a specific level of detail. That approach is adequate to achieve our purpose.  
  
  \section{Our Proposal}
  
  
  We propose a technique for generating multi-resolution terrains using the PGP. The terrain generated from out method is going to have global features. While zoomed in, not only its geometry is going to be refined, but details are going to be introduced through fractal noise usage. To achieve that we are going to further analyze state-art methods and their algorithms. 
  
  Multi-resolution terrain rendering with TS are going to be further analyzed. We are also going to do the same feat about PCG techniques suitable for generating terrains and details. Once that analysis is completed an adequate data structure for our proposed method must be either proposed or implemented. Then we are going to generate a functional prototype to validate the proposed technique.
  
  Our proposed technique must be coupled with the fact that planetary objects have specific ways to threat with model precision, view accuracy, horizon curvature representation, data dimensionality \cite{cozzi2011}. Planetary renders usually deals with atmosphere representation. Our proposal does not focus on real-time atmosphere rendering, a basic implementation is going to be used \cite{cozzi2011}. The terrains generated with our proposed method must be comprised of high geometric details, its render must support and mix GPU techniques for refinement, subdivision and simplification, thus being capable of generating and rendering multi-resolution terrains with at least some realistic appeal.
  
  The algorithm of our proposal is subdivided into two main steps, the first part runs primarily on the CPU, while the later is executed on the GPU. It is going to be comprised of the following steps: On the CPU  \begin{enumerate*}[label=(\textit{\roman*})]
    \item a low quality planetary base grid mesh is going to be instantiated,
    \item the base grid mesh is going to be associated to a control mesh data structure. Then on the GPU
    \item a procedural heightmap is going to be generated
    \item the base grid mesh is going to be displaced accordingly to the heightmap in the Vertex Shader,
    \item the base grid mesh geometry is going to be refined with TS based on visibility,
    \item the terrain is rasterized,
    \item once we have the terrain curvature, in the FS, lighting and color blending are going to be applied,
    \item the terrain is rendered,
    \item through user interaction the LOD is going to be dynamically applied,
    \begin{enumerate*}[label=(\textit{vi.\roman*})]
      \item on another pass in the FS, if the terrain is zoomed in, details are going to be further added through the introduction of tiling refinement with fractal noise and displacement \cite{bangay2017},
      \item if the terrain is zoomed out, the process renders a lower quality LOD mesh using the TS.
    \end{enumerate*}
  \end{enumerate*} The Figure \ref{fig:teaser} depicts our proposal's pipeline for procedurally generating planets and multi-resolution terrains.
  
  Our proposal aims to use methods which doesn't require or deal with any pre-generated data. Our proposal does not want to only mix two distinct techniques, but we aim to turn terrain rendering with PCG more efficient and concise, fully exploring current hardware rendering capability, under the perspective of efficiency in algorithms and routines. Texturing terrains is not a trivial task, specially for spherical surfaces. 
  
  UV mapping comes as the de-facto way to texturize a terrain through polar coordinates. However using these approaches lead to incoherences. Planets are not perfect spheres, they tend to be oblate, thus using polar coordinates may show distortions on the poles and along the equator. The terrain textures we produce are going to be generated in the Shaders, it allows us to take advantage of three planar texturing, which is best suit for round objects.
  
  In the context of this proposal techniques for CPU-GPU coupled computation \cite{bernhardt2011} and planetary scale composition \cite{kooima2009} apply. They both guide exactly on what is required towards dealing with large data and specific configurations for planetary bodies. Due to its simplicity on dealing issues that arise with rendering terrains with TS, \cite{ge2017} is going to be the default way to manage LOD and tessellation. We are going to use both techniques proposed by \cite{boubekeur2005} and \cite{bangay2017} to add details to the highest LOD possible through noise and displacement mapping.
  
  It is worth to emphasize that high quality geometry generated with our proposed method resides on the GPU. Thus General Purpose GPU -- GPGPU programming may be necessary to calculate collisions, to store a mesh or to retrieve the geometry for CPU processing. We are going to implement functional prototypes from our proposed method. We also are going to release our proposed method as a framework which can be integrated into game engines such as Unity or Unreal. We further aim to share its code to the open-source and scientific community for reviews and further collaboration.
  
  \section{Conclusion}
  
  We presented our proposal for a multi-resolution technique terrain for games. We aim to make it possible combining two strategies with CPU large scale terrain rendering, and with GPU precise low scale terrain rendering. We are going to use PCG to both generate those terrains and to add details on the highest possible LOD. Those details are going to be added on the FS. We are also going to use TS to manage and generate LOD on those terrains. To the best of our knowledge there is an existent technique which unified both the global and local aspects for rendering large terrains. Our next steps comprise of studying terrain rendering with TS implementations and adapt them to usage with PCG. We are also going to face challenges related to planetary terrains. Once we overcome those challenges we intend to release a functional prototype to evaluate our proposal, we also inted to share our technique as a library which may be integrated on known game engines. As technical improvement we may research on using geometry shader to balance LOD control and GPGPU for refined mesh retrieval.
  
  \bibliographystyle{plainurl}
  \bibliography{biblio}
\end{document}